\documentclass[aps,twocolumn]{revtex4}
\usepackage{graphicx}

\def\etal{\emph{et al. }}
\def\insitu{\emph{in situ }}

\begin{document}

\title{Local thermometry technique based on proximity-coupled superconductor/normal-metal/superconductor devices}
\author{Z. Jiang,$^1$ H. Lim,$^1$ J. Eom$^2$ and V. Chandrasekhar,$^1$}
\affiliation{$^1$Department of Physics and Astronomy, Northwestern University, Evanston, IL 60208, USA}
\affiliation{$^2$Department of Physics, Sejong University, Seoul 143-747, Korea}
\date{\today}

\begin{abstract}
In mesoscopic superconductor/normal-metal/superconductor (SNS) heterostructures, it is known that the resistance of the normal metal between the superconductors has a strong temperature dependence. Based on this phenomenon, we have developed a new type of thermometer, which dramatically enhances our ability to measure the local electron temperature $T_e$ at low temperatures. Using this technique, we have been able to measure small temperature gradients across a micron-size sample, opening up the possibility of quantitatively measuring the thermal properties of mesoscopic devices.
\end{abstract}

\maketitle

Interest has grown in recent years in the thermal properties of mesoscopic samples, after the electrical characteristics of such devices have been extensively studied in the past two decades. In order to make thermal and thermoelectric measurements, a temperature gradient needs to be set up across the sample. Experimentally, this can be done by applying a direct (dc) current through a metallic wire to heat up a part of the sample to a temperature higher than the substrate temperature $T_b$ \cite{dikin,parsons}, while keeping the substrate and other parts of the sample cold. The length of this heater line is normally much longer than the electron-electron scattering length $L_{ee}$, but shorter than the electron-phonon scattering length $L_{ep}$. In this configuration, the electrons in the center of the heater achieve local thermodynamic equilibrium by energy exchange between electrons \cite{henny1}. As the local electron temperature $T_e$ in the heater is higher than $T_b$, a temperature gradient can be generated in the electron bath. Conventional low temperature thermometers cannot be used to make a direct measurement of $T_e$ on mesoscopic samples, because their physical size is much larger than the sample of interest. Hence special thermometry techniques are needed on the submicron length scale.

Heretofore, only a few techniques have been used to measure $T_e$ directly. Aumentado \etal introduced a thermometer which could determine the electron temperature over $\sim$100 nm size scale \cite{joe}. This thermometer was based on the proximity effect, the fact that the resistance of a normal metal wire in proximity to a superconductor shows a measurable temperature dependence at temperatures below the critical temperature, $T_c$, of the superconductor. From the base temperature of dilution refrigerator to $T$$\sim$0.8 K, the overall resistance change of such a thermometer is normally $\sim$1\% of the normal state resistance. Unfortunately, a $1\%$ resistance change cannot always give one enough sensitivity to precisely measure a small temperature difference across a sample. Another technique to directly determine $T_e$ is through noise thermometry, since the Johnson-Nyquist noise of the electrons, defined by $S_V$=$4kTR$, has a linear temperature dependence. However, this technique imposes restrictive constraints on the design of the samples. In addition, noise measurements typically have a cutoff sensitivity below which the measurement accuracy drops. For example, Henny \etal mention a cutoff at a value of $2$$\times$$10^{-20}$ V$^{2}$s \cite{henny2}, which requires a minimum thermometer resistance of 3.6 k$\Omega$ at 100 mK. Superconducting quantum interference devices (SQUIDs) can increase the sensitivity of noise measurements, but need relatively complicated circuits and sample fabrication \cite{steinbach,jehl}.

In this paper, we describe a more sensitive thermometer based on the resistance of a four-terminal Al/Au device that shows a strong temperature dependence at low temperatures. The magnitude of the thermometer resistance change, from 18 mK to $\sim$350 mK, is up to 102$\%$ of the normal resistance measured at a temperature just above $T_c$. Above 350 mK, the thermometer enters the proximity effect regime and can be used like the thermometer described in \cite{joe}.

\begin{figure}[b]
\includegraphics[height=5.6cm]{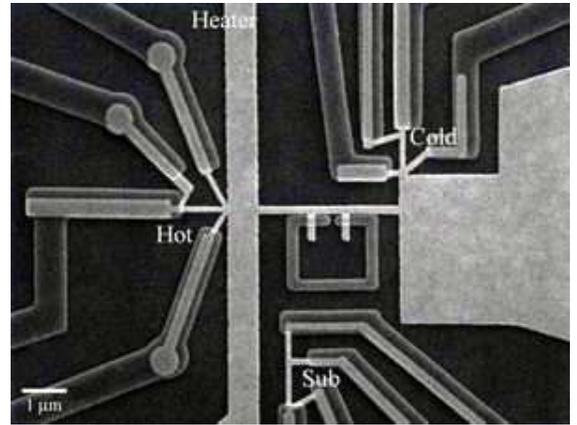}
\caption{Scanning electron microscopy image of the device structure. The brighter regions are composed of normal metal (Au), while the darker regions are superconductor (Al). The device consists of three thermometers. The one coupled to the heater is labeled ``Hot'' (length: 0.95 $\mu$m, width: 0.13 $\mu$m); the one away from the heater on the right side is labeled ``Cold'' (length: 0.67 $\mu$m, width: 0.13 $\mu$m); the one on the substrate is labeled ``Sub'' (length: 0.91 $\mu$m, width: 0.12 $\mu$m).}
\label{fig1}
\end{figure}

The device discussed in this paper was patterned onto an oxidized Si substrate (100-nm-thick) using conventional multi-level electron beam lithography. Figure 1 shows a scanning electron microscopy (SEM) image of our device. This device was fabricated for thermal conductance measurements on Andreev interferometers; these measurements will be described in detail elsewhere. Here we concentrate on the design and fabrication of the thermometers used in the experiments. The brighter regions in the image of Fig. 1 are a 46-nm-thick Au film, which was deposited first. A 76-nm-thick Al film (darker regions) was deposited on top of the Au in another level of lithography, after an \insitu Ar plasma etch was used to clean the Au surfaces in order to obtain good normal-metal/superconductor (NS) interfaces. The device is divided into four sections (see Fig. 1): (a) The ``Hot'' section, on the left, consists of a 0.73-$\mu$m-wide long heater line and a ``Hot'' thermometer. Applying a dc current through the heater increases the electron temperature $T_e$ above the substrate temperature $T_b$. (b) The ``Cold'' section, on the right, includes a large normal metal pad and a ``Cold'' thermometer. (c) The sample section consists of the structures between the ``Hot'' and ``Cold'' sections. The sample is an Andreev interferometer, which is a 3.18-$\mu$m-long and 0.18-$\mu$m-wide Au wire with an Al loop hooked up in the middle. Since the ``Hot'' and the ``Cold'' sections are at different temperatures, a temperature gradient is generated across the sample. The temperature difference can be accurately controlled by the dc current through the heater line. (d) The last section is a single thermometer electrically isolated from the rest of the sample, which measures the temperature of the substrate. This ``Sub'' thermometer can be used to monitor the heat leak from the heater to the substrate by electron-phonon scattering.

\begin{figure}[b]
\includegraphics[width=7.5cm]{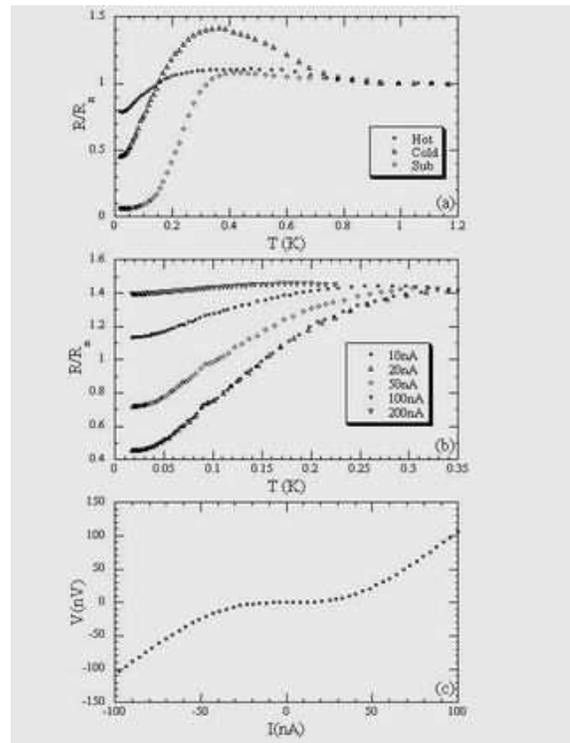}
\caption{(a) Normalized resistance of the ``Hot'', ``Cold'' and ``Sub'' thermometers as a function of temperature. The normal state resistance $R_n$ of these three thermometers are 4.08$\Omega$, 4.19$\Omega$ and 5.28$\Omega$ respectively. (b) Temperature-dependent resistance of the ``Cold'' thermometer at different measurement ac currents. The suppression of the resistance change at higher ac currents indicates Josephson coupling between superconducting leads. (c) $I$-$V$ curve of the ``Cold'' thermometer at the substrate temperature $T_b$=17.6 mK after subtracting the $I$=0 value. Extrapolation of the high current part of the curves gives a `critical current' of $I_c$$\simeq$40 nA.}
\label{fig2}
\end{figure}

All three thermometers on the device have similar structure. They consist of a short normal metal wire and four probes, each covered by a superconducting lead. The superconducting contacts are designed to be close to each other to increase the coupling between them, but not close enough to cause a supercurrent to flow between them, as this would effectively short the resistance of the normal metal wire, and defeat its use as a thermometer. The superconducting contacts also drastically reduce the heat flow through the leads, giving a relatively uniform electron temperature profile across the length of each thermometer. Figure 2(a) demonstrates the normalized resistance of all three thermometers as a function of temperature. The resistance was measured by a conventional four-probe technique using an ac resistance bridge with a 20 nA excitation current. In order to increase the measurement sensitivity, a step-up transformer, an instrumentation amplifier (AD624) and a lock-in amplifier (PAR 124) have been used in series to amplify the signal out of the resistance bridge. The instrumentation amplifiers, current source and resistance bridges are battery powered, and placed as close as possible to the top of the dilution refrigerator in a mumetal shielded box to reduce interference and inadvertent heating from line frequency sources. In addition, each electrical line into the dilution refrigerator is filtered with a $\pi$-filter with a cut-off frequency of 5 MHz in order to minimize sample heating due to ambient radio-frequency (rf) sources. The background noise from the spectrum of the signal from the samples using this setup, as measured by a spectrum analyzer attached to the lock-in amplifier, was approximately 6 nV/$\sqrt{\text{Hz}}$, close to the 4 nV/$\sqrt{\text{Hz}}$ expected input noise of the AD624, and showed an essentially flat frequency response (with no peaks at the harmonics of the line frequency) up to the maximum frequency of 100 kHz of the analyzer. In order to maximize the gain of the transformer, a frequency of 103 Hz was used for the measurements.

The behavior of the thermometer resistance below $T_c$ can be divided into two regimes. The first regime is the low temperature regime, which extends from the base temperature of the dilution refrigerator up to temperatures of 300-350 mK. The thermometer shows a strong temperature dependence in this regime, much larger than can be expected from the enhancement of conductance due to the proximity effect.  It appears that this strong resistance change is associated with Josephson coupling between the pairs of superconducting contacts on either end of the thermometers. Due to this coupling, the resistance of those parts of the thermometers near these contacts drops significantly, but the entire thermometer does not go superconducting, since distance between superconducting contacts on opposite ends of thermometer is too large for significant Josephson coupling to occur between them in our measurement temperature range.  A number of experimental facts support this picture.  First, the temperature range over which this strong temperature dependence is observed correlates well with the temperature range where Josephson coupling between the superconducting contacts is expected. The temperature at which one expects to see Josephson coupling between the superconducting contacts is determined by the correlation energy $E_c$=$\hbar D/L^2$ \cite{courtois}, where $D$ is the electronic diffusion coefficient in the normal metal, and $L$ the length between the superconducting contacts. Taking, as an example, the ``Cold'' thermometer, with the measured value of $D$=104 cm$^2$/sec, and the length $L$=0.485 $\mu$m between the nearest pair of contacts, the temperature below which Josephson coupling between the two superconducting contacts should become significant is $E_c/k_B$$\simeq$336 mK \cite{courtois}, which is in good agreement with the experimentally observed value of $\sim$350 mK where the resistance starts to drop rapidly (see Fig. 2(a)). Second, the resistance of the thermometers in this regime is strongly dependent on the measurement current. Figure 2(b) illustrates the temperature-dependent resistance $R(T)$ at low temperatures of the ``Cold'' thermometer taken with different values of the ac measurement current. As the ac measurement current is reduced from 200 nA to 20 nA, the overall change in resistance from approximately 350 mK to base temperature increases substantially.  Below 20 nA, there is essentially no change in the resistance of the thermometer. Third, the differential resistance $dV/dI$ of the thermometers as a function of the dc current $I$ through the thermometer is strongly reminiscent of the differential resistance of a SNS junction, except that the differential resistance at $I$=0 is not zero.  Figure 2(c) shows the current-voltage ($I$-$V$) curve of the ``Cold'' thermometer, obtained by integrating the $dV/dI$ vs. $I$ curve after the $I$=0 value of the differential resistance was subtracted. The overall shape of this $I$-$V$ characteristic is very much like the $I$-$V$ characteristic for a SNS junction. Extrapolation of the high current part of the curves gives a `critical current' of $I_c$$\simeq$40 nA. However, the $I$=0 resistance of the thermometer is not 0.  The resistance of the thermometer would vanish only if the Josephson coupling between superconducting contacts on opposite sides of the thermometer is appreciable.  Using $L$=1.15 $\mu$m as the length between superconducting contacts on opposite sides of the ``Cold'' thermometer, the temperature below which Josephson coupling would be appreciable is $E_c/k_B$$\simeq$60 mK, at the lower end of our temperature range.  Consequently, the critical current between the superconducting contacts on opposite sides of the thermometer is expected to be exponentially smaller, so it is not surprising that we observe a finite resistance of the thermometer even at our lowest temperatures.  It should also be noted that, in spite of our best efforts to shield all sources of extrinsic noise, there may still be some noise coupled to the sample, which suppresses the critical current across the thermometers.   

The second regime in Fig. 2(a) is the proximity effect regime (from $\sim$350 mK up to $T_c$). The resistance of a normal metal in proximity to a superconductor should decrease below its normal state value $R_n$ when the temperature drops below $T_c$ of the superconductor. This is the so-called proximity effect. It is due to the penetration of the order parameter of superconductor into the normal metal \cite{gennes}. However, Fig. 2(a) shows an \textit{increase} in resistance of all the thermometers in this temperature regime. Similar behavior has been observed before in proximity coupled devices \cite{wilhelm}, and was explained as arising from current redistribution in samples with a four-probe configuration, where the width of the sample is of the same order as its length, as the sample is cooled below $T_c$. In our device, the ``Cold'' thermometer, which has the largest width, but the shortest length, is the most likely candidate for this mechanism. This correlates with the fact that the ``Cold'' thermometer has the strongest temperature dependence of the three thermometers just below $T_c$. 

\begin{figure}[b]
\includegraphics[width=7.9cm]{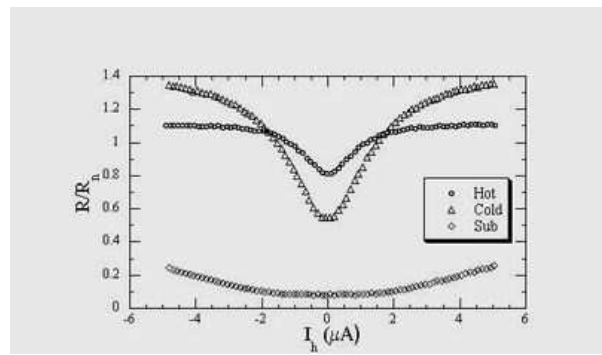}
\caption{Normalized resistance of the three thermometers as a function of dc current through the heater line at the substrate temperature $T_b$=49.5 mK.}
\label{fig3}
\end{figure}

\begin{figure}[t]
\includegraphics[width=7.9cm]{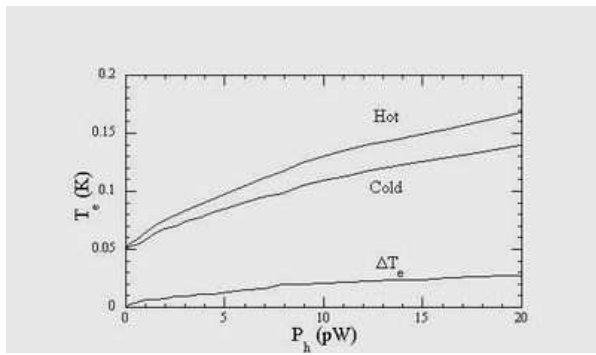}
\caption{Local electron temperature $T_e$ of the ``Hot'' and ``Cold'' thermometers and the temperature gradient $\Delta T_e$ across the sample as a function of heater power $P_h$ at the substrate temperature $T_b$=49.5 mK. This figure is obtained from the $R(T)$ and $R(I_h)$ data shown in Fig. 2(a) and Fig. 3.}
\label{fig4}
\end{figure}

The large change in the resistance of these devices makes them excellent candidates for use as local electron thermometers in mesoscopic thermal transport experiments. For this purpose, the thermometers are calibrated at a fixed base temperature $T_b$, as a function of the dc current $I_h$ through the heater \cite{joe}. Figure 3 shows the result of this measurement, at $T_b$=49.5 mK. These measurements are then numerically cross-correlated with the measured $R(T)$ curves shown in Fig. 2(a), to obtain the electronic temperature $T_e$ as a function of $I_h$. The resulting $T_e$ for the ``Hot'' and ``Cold'' thermometer is shown in Fig. 4, expressed as a function of the power through the heater $P_h$=$I_h^2 R_h$. Figure 4 shows that a significant fraction of the power generated in the heater flows through the sample, because $T_e$ in both thermometers increases substantially when we increase the heater power, while the electron temperature differential $\Delta T_e$ increases only gradually. Furthermore, note that the substrate thermometer resistance shown in Fig. 3 has a very weak dependence on $I_h$ at low values of $I_h$. This demonstrates that there is very little heat leak through the substrate at low temperatures and small heater currents, showing that heat loss from the electron bath due to electron-phonon interactions is very small in this regime.

The ability to measure small temperature differences in the electron temperature opens up the possibility of quantitatively measuring the thermal properties of mesoscopic samples. For example, in the sample of Fig. 1, if we have a way of determining the heat current through the Andreev interferometer, we can directly measure its thermal conductance. In this sample, the electrical connections from the heater line and thermometers extending out to the outside patterns are all designed to be superconducting. The advantage of these superconducting leads is that they block the heat transport from the device to its surroundings. At temperatures well below $T_c$, the thermal conductance of the superconductors is so small that we can assume most of the power $P_h$ generated in the heater flows out only through the Andreev interferometer. (Heat conduction due to phonons is also negligible below 200 mK, as noted above.) The heat flow through the Andreev interferometer can therefore be determined by simply measuring $P_h$, which is represented by the abscissa in Fig. 4. The thermal conductance is then $G_T$=$P_h/\Delta T_e$. For the data represented in Fig. 4, we obtain $G_T$=0.12 nW/K, at a heater power of 1 pW.  

In summary, we have developed a new type of thermometer, which has a simple structure but strong temperature dependence. It can be used to quantitatively measure small temperature gradients across mesoscopic samples, and to explore thermal properties on the micron and nanometer scale.

This work is supported by the NSF through grant number DMR-0201530, and by the Korea Research Foundation Grant (KRF-2001-003-D00051).


\begin{thebibliography}{text}
\bibitem{dikin}D.A. Dikin, S. Jung, and V. Chandrasekhar, Phys. Rev. B {\bf 65}, 12511 (2002).
\bibitem{parsons}A. Parsons, I.A. Sosnin, and V.T. Petrashov, cond-mat/0107144 (July 11, 2002).
\bibitem{henny1}M. Henny, S. Oberholzer, C. Strunk, and C. Sch\"{o}nenberger, Phys. Rev. B {\bf 59}, 2871 (1999).
\bibitem{joe}J. Aumentado, V. Chandrasekhar, J. Eom, P.M. Baldo, and L.E. Rehn, Appl. Phys. Lett. {\bf 75}, 3554 (1999).
\bibitem{henny2}M. Henny, H. Birk, R. Huber, C. Strunk, A. Bachtold, M. Kr\"{u}ger, and C. Sch\"{o}nenberger, Appl. Phys. Lett. {\bf 71}, 773 (1997).
\bibitem{steinbach}A.H. Steinbach, J.M. Martinis, and M.H. Devoret, Phys. Rev. Lett. {\bf 76}, 3806 (1996).
\bibitem{jehl}X. Jehl, P. Payet-Burin, C. Baraduc, R. Calemczuk, and M. Sanquer, Phys. Rev. Lett. {\bf 83}, 1660 (1999).
\bibitem{courtois}H. Courtois, Ph. Gandit, D. Mailly, and B. Pannetier, Phys. Rev. Lett. {\bf 76}, 130 (1996).
\bibitem{gennes}P.G. de Gennes, \emph{Superconductivity of Metals and Alloys}, (Benjamin, New York, 1964).
\bibitem{wilhelm}F.K. Wilhelm, A.D. Zaikin, and H. Courtois, Phys. Rev. Lett. {\bf 80}, 4289 (1998).
\end{thebibliography}
\end{document}